\documentclass{cimento}

\usepackage{graphicx}  

\usepackage[utf8]{inputenc}
\usepackage[T1]{fontenc}
\usepackage{subfig}
\usepackage{ulem}

\title{Ab initio Circular Dichroism with the Yambo code: beyond the Independent Particle approximation}
\author{E.~Molteni\from{ins:x}\thanks{elena.molteni@mlib.ism.cnr.it}\ETC,
G.~Mattioli\from{ins:x},
        \atque
D.~Sangalli\from{ins:x}
}
\instlist{\inst{ins:x}   Istituto di Struttura della Materia - CNR (ISM-CNR), Division of Ultrafast Processes in Materials (FLASHit), Area della Ricerca di Roma 1, Via Salaria km 29.300, CP 10, Monterotondo Scalo, Roma, Italy}

\begin{document}

\maketitle

\begin{abstract}
Circular dichroism (CD) spectroscopy is a useful technique for characterizing chiral molecules. It is more sensitive than total absorption to molecule conformation, and it is routinely used to identify enantiomers. 
We present here absorption and CD spectra within the Time Dependent (TD) B3LYP approximation in c-GlyPhe, a cyclo-dipeptide containing an aromatic group. Results from codes in localized basis-set (Orca and MolGW) are carefully compared with the novel TD-B3LYP implementation we developed in the Yambo code, that uses a plane-wave basis set. 
\end{abstract}

\section{Introduction}

Cyclo-dipeptides (CDPs) are chiral molecules of interest both from the biological and pharmacological point of view~\cite{Mishra_Molecules_2017}, and are possible building blocks for nanodevices~\cite{Zhao_PepSci_2020,Jeziorna_CrystGrowthDes_2015}.
Both in the gas phase and in solution they can present different geometric structures (conformations or enantiomers). Circular dichroism (CD) spectroscopy, due to the different absorption of left {\it vs} right circularly polarized light by chiral systems, is a powerful tool routinely used to identify enantiomers of chiral molecules.

In previous works we studied the electronic\cite{Molteni_PCCPdipep2021} and optical properties\cite{molteni2022ab}, both absorption and CD, of three CDPs containing aromatic groups, i.e. c-GlyPhe, c-TrpTyr and
c-TrpTrp. We explored in great detail the electronic levels, by comparing different levels of approximation, and also focusing on their conformational dependence. For the optical properties instead we mostly focused on the selection rules between two electronic levels and at the independent-particles (IP) level.
However, a more accurate modelling of absorption and CD requires the introduction of electronic correlation in the excited state calculations. In this work we compute optical absorption and CD within Time Dependent Density Functional Theory (TDDFT) for the lowest energy conformer of c-GlyPhe dipeptide. 

Most of the available codes for computing the optical properties of molecules are based on Gaussian-type (GT) localized basis sets. Here we use the the MolGW\cite{molgw2016} and Orca\cite{Orca_general} codes as a reference. Moreover we show results obtained with the QuantumEspresso (QE)\cite{QE_2017} 
and Yambo\cite{Sangalli2019} codes using a plane-wave (PW) basis set. 
TDDFT absorption spectra are also compared with experimental data, showing that corrections beyond IP are not negligible. 

\section{Results}\label{sec:results}

\begin{figure}[h]
\begin{center}
  \includegraphics[width=.53\linewidth]{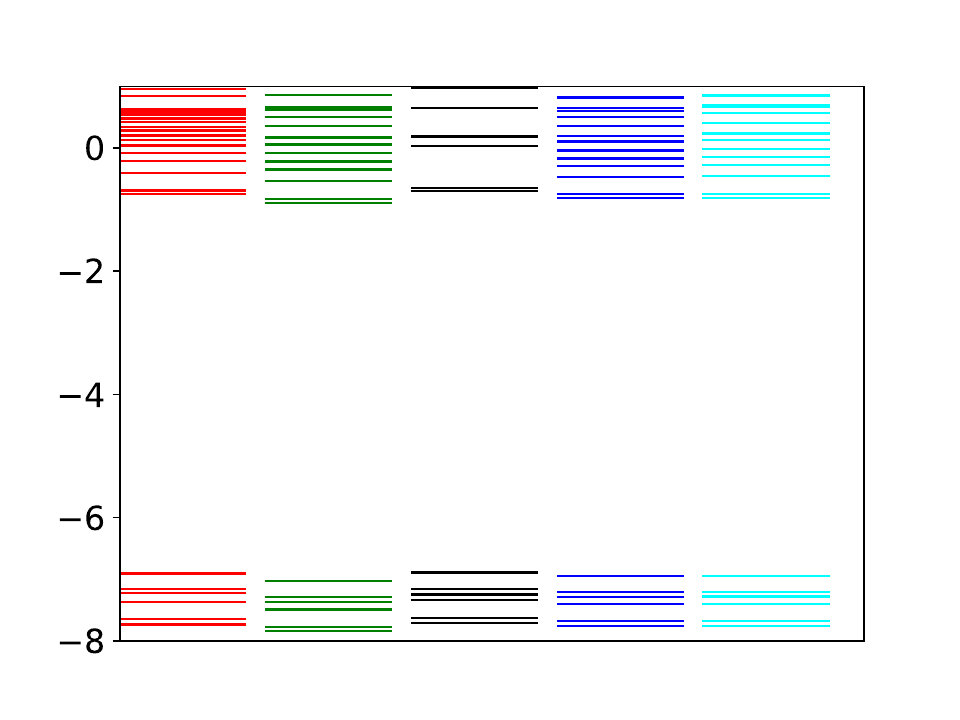}
  \includegraphics[width=.46\linewidth]{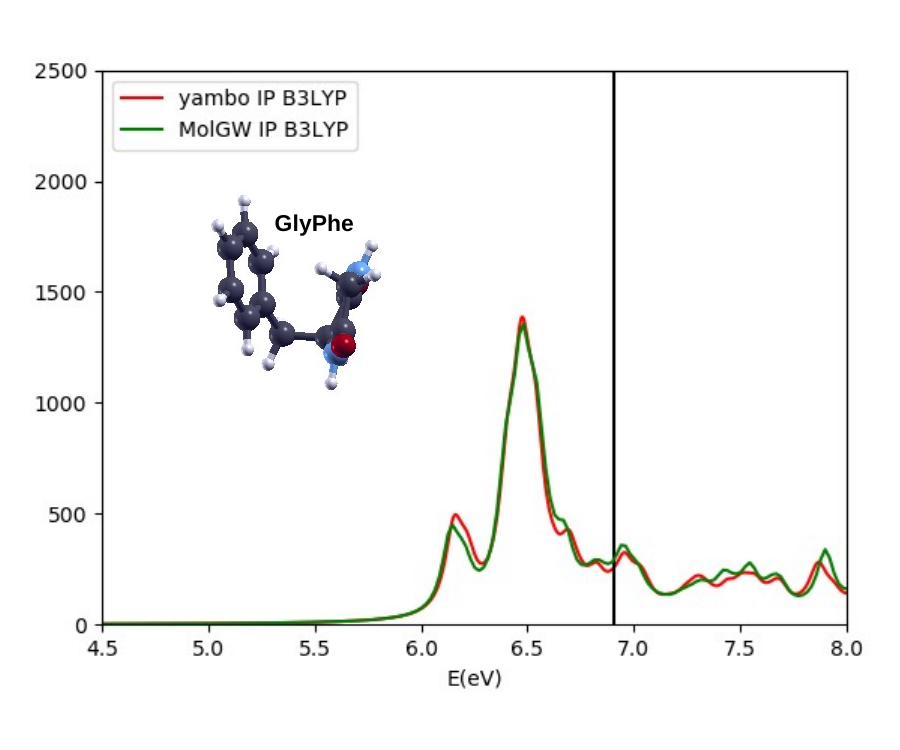}
\end{center}
\caption{Left panel: B3LYP DFT electronic levels of c-GlyPhe obtained with: QE (red), MolGW (green), Orca with def2-QZVPP basis and def2/J auxiliary basis (black), Orca with aug-cc-pVQZ basis (blue) or ma-def2-QZVPP basis (cyan), both with aug-cc-pVQZ/JK auxiliary basis: the vertical axis reports energies in eV.
Right panel: Absorption spectra of c-GlyPhe obtained within IP B3LYP, using: the Yambo code (red), the MolGW code (green), and geometry of the investigated molecule. A vertical black solid line indicates the $E_{vac}-E_{HOMO}$ value from DFT B3LYP QE calculations.}
\label{fig:Abs_IP-b3lyp_3dipep}
\end{figure}

We first compare (Fig.~\ref{fig:Abs_IP-b3lyp_3dipep}, left panel) the electronic energy levels of c-GlyPhe, obtained within DFT B3LYP using either the PW based QE code\cite{QE_2017} 
(red dataset), or the MolGW\cite{molgw2016} (green dataset) and Orca\cite{Orca_general} (black, blue and cyan) codes, based on GT basis. The energy distributions of occupied electronic states obtained with the two chosen computational schemes (PW {\it vs.} GT bases) are in agreement with each other, apart from possible small rigid shifts. As for empty states, PW codes (in our case QE, red levels) can capture continuum delocalized electronic states\cite{Molteni_PCCPdipep2021}. This results in an overall higher density of states in the positive energy range with respect to GT basis codes. In codes with GT bases (MolGW, Orca), augmented basis sets containing diffuse functions (green, blue and cyan datasets in Fig.\ref{fig:Abs_IP-b3lyp_3dipep} left panel) are needed in order to obtain bound empty levels in agreement with QE ones.  
The very good agreement between the B3LYP electronic levels is maintained also in the corresponding B3LYP IP absorption spectra (Fig.~\ref{fig:Abs_IP-b3lyp_3dipep}, right panel). Here we compare results from Yambo (red curves) with MolGW (green curves). We used a modified version of MolGW to print the IP spectra (which cannot be computed with Orca).

\begin{figure}
\begin{center}
  \includegraphics[width=.51\linewidth]{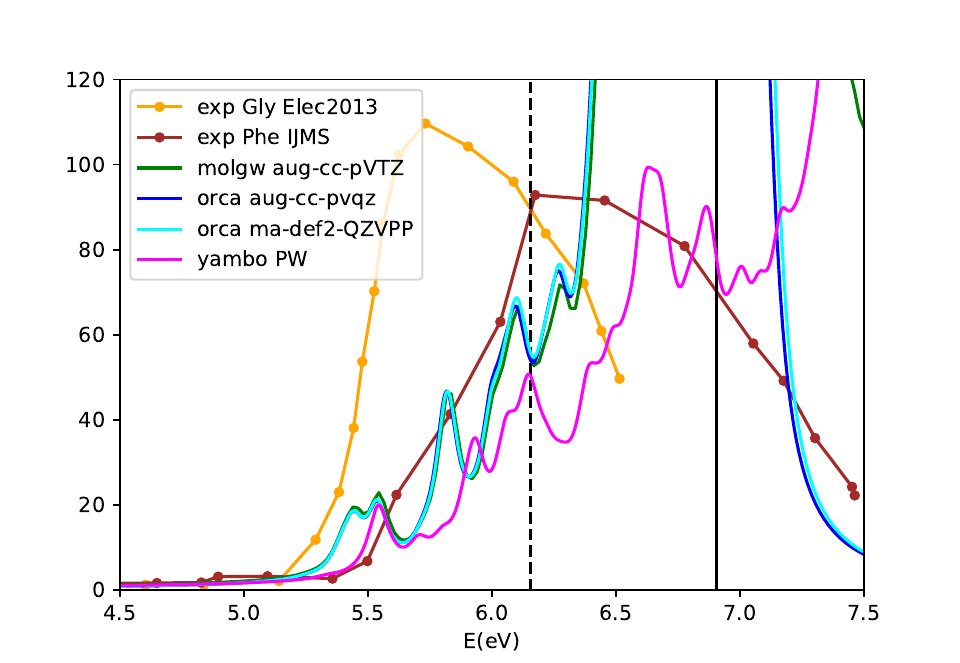}
  \includegraphics[width=.47\linewidth]{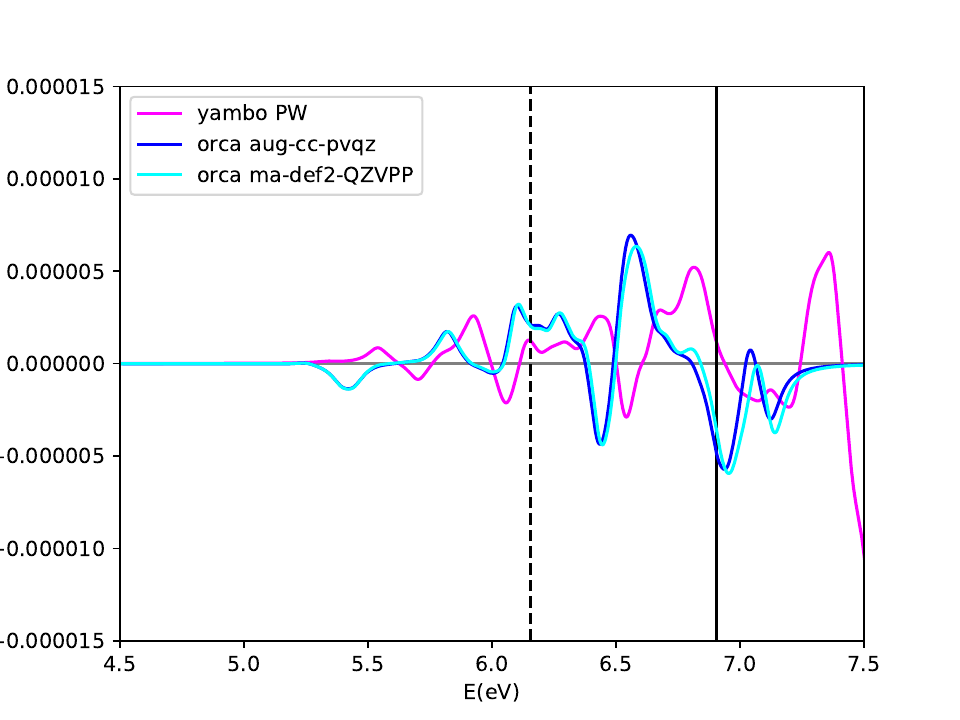}  
\end{center}
\caption{Absorption (left) and CD (right) spectra of c-GlyPhe obtained within TD-B3LYP using the Yambo code (magenta curves), or the Orca code with the ma-def2-QZVPP basis sets (cyan curves) or with the aug-cc-pVQZ basis set (blue curves) and MolGW with the aug-cc-pVTZ basis set (green curve). Experimental absorption spectra from the literature (``exp Gly Elec2013'' dataset\cite{Electr_2013}, ``exp Phe IJMS'' dataset\cite{IJMS_2009}) are reported as dotted lines. A vertical solid (dashed) black line indicates the $E_{vac}-E_{HOMO}=6.9$~eV ($E_{LUMO}-E_{HOMO}=6.15$~eV) from DFT B3LYP QE calculations.
The B3LYP vertical ionization potential is instead at 8.57~eV}
\label{fig:AbsCD_TDb3lyp_3dipep}
\end{figure}

We then move to the calculation of the TD-B3LYP spectra (Fig.\ref{fig:AbsCD_TDb3lyp_3dipep}), with Yambo, starting from QE KS B3LYP wavefunctions (magenta curves), with Orca using either the ma-def2-QZVPP or aug-cc-pVQZ basis sets (blue and cyan curves) and with MolGW using the aug-cc-pVTZ basis set (green curve). Spectra calculated with both the chosen GT codes (Orca and MolGW), using basis sets containing diffuse functions, give almost identical results. On the other hand, the results obtained with QE+Yambo are quite different from the Orca ones. Besides the different basis set (PW {\it vs.} GT), one of the reasons is that in the TD-B3LYP implementation of the Yambo code we neglect the GGA part of the $f^{xc}$ kernel. 
We have extended the TDDFT implementation of the Yambo code in three directions. (i) A TDDFT simulation now prints in output also the CD spectra. (ii) The non local part, coming from the exchange fraction, is added to the local part taking advantage of the Fock term already present in the Yambo code. (iii) We improved the computation of the matrix elements of the local part of $f^{xc}$. The real space integrals are now evaluated only in regions of space where the electronic density overcomes a threshold, $\varepsilon=1.e-7$ which can be also controlled in input. 
Thanks to (iii) the simulation time for the TDDFT step with Yambo is not longer than the time required by MolGW or Orca. By increasing $\varepsilon=1.e-6$ Yambo is even faster and less memory consuming than MolGW and Orca, without significant changes in the computed spectra.

We now discuss more in details the absorption spectrum in Fig.~\ref{fig:AbsCD_TDb3lyp_3dipep}, left panel.
At the B3LYP level, photons with energy below 8.57 eV (vertical ionization potential of c-GlyPhe from total energy differences~\cite{Molteni_PCCPdipep2021}) can only access neutral excitations. However, we notice that a continuum of transitions, from the HOMO to vacuum levels, enter the TD-B3LYP hamiltonian starting from 6.9~eV (continuous line). The HOMO-LUMO gap gives a minimum excitation energy of 6.15~eV (dashed line; see also Fig.~\ref{fig:AbsCD_TDb3lyp_3dipep}, left panel). Excited states correlation lower the onset of the of absorption spectra to 5.4~eV, in better agreement with available experimental data (absorption spectra of the single amino acids Gly and Phe). Moreover, among experimental spectra reported in the literature there is a rather large variability in the energy position of the main peaks, depending on the experimental conditions ({\it e.g.} amino acids in different solvents, or amino acid films deposited on glass surfaces, etc.).

The main low energy TD-B3LYP absorption features are slightly red-shifted in the Orca datasets with respect to the Yambo results. A similar trend can be noticed in CD spectra too (Fig.\ref{fig:AbsCD_TDb3lyp_3dipep}, right panel), once identified the CD peaks in the corresponding absorption spectrum. Without this information, indeed, the possible variability in sign (in addition to position and intensity) of CD peaks would hinder an unambiguous assignment of CD features obtained with the  PW {\it vs.} GT computational schemes. 

We are currently working on including the GGA corrections in the Yambo code implementation. Moreover we plan to study the absorption spectra of molecules with Yambo also beyond TDDFT, via the GW+BSE scheme, and to include solvation effects to provide a closer comparison with dilute-solution measurements. 

\acknowledgments
The present work was performed in the framework the PRIN
20173B72NB research project “Predicting and controlling the fate of biomolecules driven by extreme-ultraviolet radiation”, which involves a combined experimental and theoretical study of electron dynamics in biomolecules with attosecond resolution.
EM acknowledges F. Bruneval for support with the MolGW code.

\nocite{*}
\bibliography{bib_cimento.bib} 

\end{document}